\documentclass[aps,pre,floats,floatfix,twocolumn]{revtex4}

\usepackage{ifthen}
\usepackage{ifpdf}
\usepackage{color}

\ifpdf
\usepackage{graphicx}
\usepackage{epstopdf}
\else
\usepackage{graphicx}
\usepackage{epsfig}
\fi

\usepackage{latexsym}
\usepackage{amsmath}
\usepackage{amssymb}
\usepackage{bm}
\usepackage{wasysym}

\newcommand{\const}{\mbox{const}}

\newcommand{\eexp}{\mbox{e}^}

\newcommand{\tbox}[1]{\mbox{\tiny #1}}

\newcommand{\amatrix}[1]{\begin{matrix} #1 \end{matrix}}

\newcommand{\be}[1]{\begin{eqnarray}\ifthenelse{#1=-1}{\nonumber}{\ifthenelse{#1=0}{}{\label{e#1}}}}
\newcommand{\ee}{\end{eqnarray}}

\newcommand{\hide}[1]{\textcolor{red}{[hidden text]}}

\begin{document} 

\title{Anomalous decay of a prepared state due to non-Ohmic coupling to the continuum}

\author{Itamar Sela$^{1}$, James Aisenberg$^{2}$, Tsampikos Kottos$^{2}$, Alex Elgart$^{3}$, Doron Cohen$^{1}$}

\affiliation{
$^1$Department of Physics, Ben-Gurion University, Beer-Sheva 84105, Israel\\
$^2$Department of Physics, Wesleyan University, Middletown, CT 06459, USA\\
$^3$Department of Mathematics, Virginia Tech, Blacksburg, VA 24061, USA}

\begin{abstract}
We study the decay of a prepared state $E_0$ into 
a continuum $\{E_k\}$ in the case of non-Ohmic models. 
This means that the coupling is $|V_{k,0}|\propto |E_k{-}E_0|^{s{-}1}$ 
with ${s\ne1}$. We find that irrespective of model details 
there is a universal generalized Wigner time~$t_0$ 
that characterizes the evolution of the survival probability $P_0(t)$. 
The generic decay behavior which is implied 
by rate equation phenomenology is a slowing down stretched exponential,   
reflecting the gradual resolution of the bandprofile.  
But depending on non-universal features of the model 
a power-law decay might take over: 
it is only for an Ohmic coupling to the continuum 
that we get a robust exponential decay that is insensitive 
to the nature of the intra-continuum couplings.
The analysis highlights the co-existence of perturbative and
non-perturbative features in the dynamics.
It turns out that there are special circumstances 
in which~$t_0$ is reflected in the spreading process 
and not only in the survival probability, 
contrary to the naive linear response theory expectation. 
\end{abstract}

\maketitle

\section{Introduction}

The time relaxation of a quantum-mechanical prepared state
into a continuum due to some residual interaction 
is of great interest in many fields of physics. 
Applications can be found in areas as diverse as 
nuclear \cite{AZ02}, 
atomic and molecular physics \cite{CDG92} 
to quantum information \cite{NC00}, 
solid-state physics \cite{PAEI94,BH91}
and quantum chaos \cite{PRSB00}. 
The most fundamental measure characterizing
the time relaxation process is the so-called survival probability~$P_0(t)$,
defined as the probability not to decay before time~$t$.

The study of $P_0(t)$ goes back to the work
of Weisskopf and Wigner \cite{WW30,wigner} regarding 
the decay of a bound state into a continuum.
They have found that~$P_0(t)$  follows an exponential 
decay ${P_0(t)=\exp(-t/t_0)}$, with rate $1/t_0$
given by the Fermi Golden Rule (FGR). 

Following Wigner, many studies have adopted
Random Matrix Theory (RMT) modeling \cite{mario,flamb,izrailev,fyodo,GACMP97}
for the investigation of~$P_0(t)$, 
highlighting the importance of the statistical properties 
of the spectrum \cite{GACMP97}, 
both in the semiclassical \cite{wls,lds,brm} 
and in the many-body context \cite{S01} 
where non-exponential decay such 
as ${P_0(t) \sim \exp(-{\sqrt t})}$ may arise.

Despite the interest in specific problems where   
deviations from the Wigner theory arise, 
a theoretical investigation of the time relaxation
for prototype RMT models is still missing, and also
the general (not model specific) perspective is lacking.
This unbalanced situation should be contrasted 
with the arena of spectral statistics \cite{haake}, 
where one can find on the one hand 
elaborated mathematical studies of generics RMT models, 
and on the other hand system specific investigations 
that explore non-universal features 
that cannot be captured by RMT.

In this paper we would like to explore the limits 
of universality in decay problems using an RMT perspective.  
Specifically we explore the decay of  
an initially prepared state into the continuum
assuming non-Ohmic rather than Ohmic circumstances.  
The latter notions are precisely defined in the next section.   
We show that the survival probability $P_0(t)=g(t/t_0)$ is characterized
by a generalized Wigner decay time $t_0$ that depends in a non-linear
way on the strength of the coupling. We also establish that the
scaling function $g$ has distinct universal and
non-universal features. It is only for Ohmic coupling to the continuum
that we get a robust exponential decay that is insensitive 
to the nature of the intra-continuum couplings.
In addition to $P_0(t)$ we investigate other characteristics of the evolving wavepacket: 
the variance $\Delta E_{\tbox{sprd}}(t)$ and the $50\%$ probability width $\Delta E_{\tbox{core}}(t)$
of the energy distribution, that describe universal and non-universal
features of its decaying component.

The structure of the paper is as follows: In Section~II 
we define the Friedrichs model (FM) \cite{friedrichs} 
and the generalized Wigner model (WM), 
and discuss the numerical approach used in the subsequent sections. 
The quantities under investigation, and their physical meaning is discussed in Section~III. 
In Section~IV we present an overview of the relevant time scales that dictate the dynamics 
of our models. In Section~V we present analytical and numerical results 
for the Local Density of States (LDOS) of the FM and of the WM. The study of 
the wavepacket dynamics is presented in Sections~VI-VIII: first we analyze 
the decay of the survival probability and later the evolution of the energy distribution. 
Our conclusions are given at the last section IX, where we also discuss 
the crossover from the universal to the non-universal behavior.

\section{Modeling}

\begin{figure}[h!]
\includegraphics[width=0.75\hsize,clip]{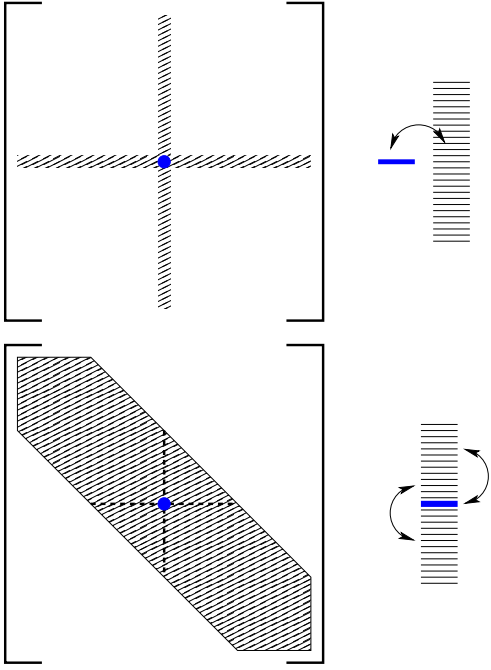}
\caption{(color online) 
Schematic illustration of the coupling matrix $V_{nm}$ 
for the FM (\textit{upper panel}) and the WM (\textit{lower panel}).
}
\label{fig:V}
\end{figure}

We analyze two models whose dynamics is generated by an Hamiltonian
\be{0}
{\cal H} = {\cal H}_0+V
\ee
with ${\mathcal{H}_0=\mbox{diag}\{E_n\}}$ and $n\in{\mathbb Z}$.
The system is prepared initially in the eigenstate
corresponding to $E_0$, and the coupling to the other levels is
characterized by the spectral function
\be{546} 
\tilde{C}(\omega)
\ = \ 
\sum_{n\neq0} 
{|V_{n,0}|^2} 2\pi\delta(\omega
-(E_n{-}E_0)) 
\ee
In the so-called Ohmic case  $\tilde{C}(\omega)$ 
is finite for ${\omega\sim0}$ and accordingly FGR 
suggests a well definite finite rate of decay.
But our interest is in so-called  non-Ohmic 
circumstances for which $\tilde{C}(\omega)$  
either vanishes or diverges as ${\omega\rightarrow0}$.   
To be specific we assume the standard form   
\be{547}
\tilde{C}(\omega)
\ = \
2\pi \epsilon^2 |\omega|^{s-1} \eexp{-|\omega|/\omega_c}
\ee
Note that $\tilde{C}(\omega)$ has $1/\mbox{time}$ units, 
and hence $\epsilon$ has $1/\mbox{time}^{2{-}s}$ units. 
In order to define the model one should specify  
the exponent~$s$, the strength of the coupling $\epsilon$, 
the density of states $\varrho$ and the bandwidth ${\omega_c=b\varrho^{-1}}$.
But this is not enough. One also has to specify the couplings 
between the other levels $E_{n\ne 0}$. Here we distinguish between 
two cases that are illustrated in Fig.(\ref{fig:V}) 
and discussed below.

{\em The Friedrichs model} (FM) \cite{friedrichs} features   
a distinguished energy level $E_0$ that is coupled 
to the rest of the levels $E_{n \neq 0}$ by a rank two matrix. 
This means that the other levels are not coupled.   
For example the following is a $6\times6$ FM matrix with ${b=2}$.  
\be{0}
{\cal H} = \left(
\begin{array}{cccccc}
E_0 & V_{01} & V_{02} & 0 & 0 & 0 \\
V_{10} & E_1 & 0 & 0 & 0 & 0 \\
V_{20} & 0 & E_2 & 0 & 0 & 0 \\
0 & 0 & 0 & E_3 & 0 & 0 \\
0 & 0 & 0 & 0 & E_4 & 0 \\
0 & 0 & 0 & 0 & 0 & E_5 \\
\end{array}\right)
\ee
In the FM case the dimensionless bandwidth~$b$  
defines the effective size of the matrix as ${N=b+1}$,   
because out-of-band levels cannot be reached.

{\em The Wigner model} (WM) \cite{wigner} features  
a perturbation matrix that does not discriminate between the levels, 
and is given by a banded {\em random} matrix. For example: 
\be{0}
{\cal H} = \left(
\begin{array}{cccccc}
E_0 & V_{01} & V_{02} & 0 & 0 & 0 \\
V_{10} & E_1 & V_{12} & V_{13} & 0 & 0 \\
V_{20} & V_{21} & E_2 & V_{23} & V_{24} & 0 \\
0 & V_{31} & V_{32} & E_3 & V_{34} & V_{35} \\
0 & 0 & V_{32} & V_{43} & E_4 & V_{45} \\
0 & 0 & 0 & V_{43} & V_{54} & E_5 \\
\end{array}\right)
\ee
Here we assume that all the levels are alike, 
and therefore Eq.(\ref{e547}) implies that 
\be{0}
\overline{|V_{n,m}|^2} \propto |E_n{-}E_m|^{s{-}1}
\ee
where the overline indicates an RMT averaging 
over realizations, which from now on will be implicit 
in the analysis of the WM case. Unlike the FM
case, here the dimension of the matrix $N$ has significance, 
because any level can be reached by high order processes.
As far as the theoretical analysis is concerned 
we assume the matrix to be of infinite size.

The assumed form Eq.(\ref{e547}) for the spectral function $\tilde{C}(\omega)$ 
is motivated by the study of various model systems 
(e.g. those of Refs.\cite{wls,lds,brm,S01}) where the   
assumption of ``flat" band-profile looks like an oversimplification. 
Thus it constitutes the natural generalization 
for the standard FM and WM. By integrating Eq.(\ref{e547}) 
over~$\omega$ we see that the perturbation $V$ 
that appears in Eq.(\ref{e546}) is bounded provided ${s>0}$.
The ${s=1}$ case is what we refer to as the Ohmic case,
for which it is well known that both models leads
to the same exponential decay for the survival probability \cite{CDG92}.
For ${s>2}$ the effect of the continuum can be handled
using $1^{\tbox{st}}$ order perturbation theory.
We focus in the ${0<s<2}$ regime and consider the
${s\neq1}$ case for which a non-linear version of the
Wigner decay problem is encountered.

\begin{figure}[h!]
\begin{center}
\includegraphics[width=0.95\hsize,clip]{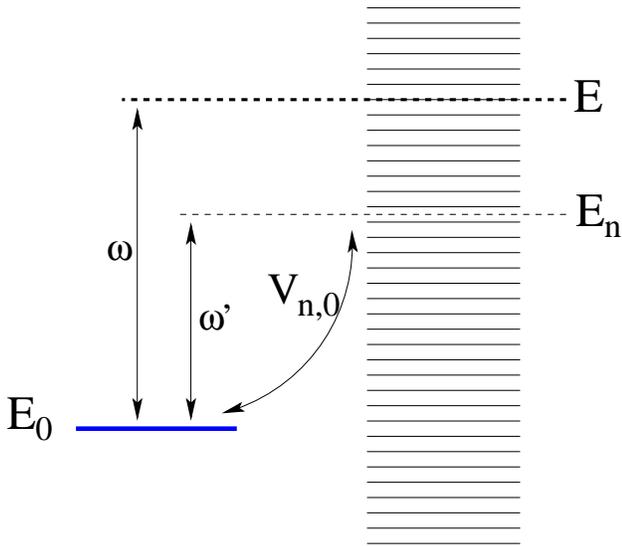}
\end{center}
\caption{(color online) 
Illustration of the energy level scheme, 
and the definition of $\omega$ and $\omega'$, 
as explained in the main text.
}
\label{fig:Ediagrame}
\end{figure}

We measure the energy taking $E_0$ as a reference
and accordingly we use the notations ${\omega \equiv E-E_0}$, 
and ${\omega_n \equiv E_n-E_0}$. 
In the continuum limit any summation 
over $\omega_n$ is replaced by an integral 
over a variable ${\omega'}$.
See Fig.(\ref{fig:Ediagrame}) for illustration.

In the numerical simulations we integrate the
Schr\"odinger equation for the amplitudes 
${\psi_n(t) = \langle n|\psi(t)\rangle}$
starting with the initial condition
$\psi_n(0)=\delta_{n,0}$. 
We use units such that $\hbar{=}1$,   
the density of states is $\varrho{=}1$, 
and $E_0{=}0$, 
and we assume a sharp bandwidth 
\be{0}
b=\varrho\omega_c  
\ \ \ \ \ \ \ \ \mbox{[bandwidth]}
\ee
The matrix elements of $V$ are taken from a Gaussian distribution 
with zero mean and variance $\epsilon^2$ in the units as defined above.
The integration is done using the self-expanding algorithm 
of \cite{wbr} to eliminate finite-size effects, 
adding $10b$ sites to each edge of the energy lattice 
whenever the probability of finding the `particle' 
at the edge sites exceeds~$10^{-12}$.

\section{Strategy of analysis}

\begin{figure}[h!]
\begin{center}
\includegraphics[width=0.95\hsize,clip]{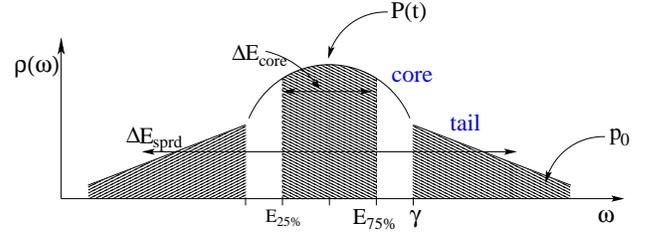}
\end{center}
\caption{(color online) 
Cartoon that illustrates the structure 
of the energy distribution $\rho(\omega)$. 
The probability that would be held by the tails 
is $p_0\ll1$ if FOPT strictly applied. It becomes 
of order unity (say 50\%) for a fully developed core-tail structure.
The core has a semicircle line shape in the WM case, 
and its border $\gamma$ is determined self-consistently.
The distribution is characterized by $P_0(t)$, 
and by $\Delta E_{\tbox{core}}$, and by $\Delta E_{\tbox{sprd}}$, 
as explained in the main text.
}
\label{fig:LDOScartoon}
\end{figure}

We denote by $|n\rangle$ the unperturbed eigenstates 
of ${\cal H}_0$ and by $|E_{\nu}\rangle$ 
the perturbed eigenstates of ${\cal H}$.
The local density of states LDOS with respect 
to the initial state $| 0 \rangle$ is defined as
\be{0}
\rho(\omega) = \sum_{\nu} |\langle E_{\nu} | 0 \rangle|^2 \delta(\omega-(E_{\nu}{-}E_0))
\ee
Starting with the initial state    
the time-dependent state $|\psi(t)\rangle$
can be represented by the amplitudes $c_n(t)$, 
such that 
\be{0}
|\psi(t)\rangle \ \ = \ \ \sum_n c_n(t) \ \eexp{-iE_nt} \ |n\rangle
\ee
The probability distribution at time $t$ is 
\be{0}
P_n(t) \ = \ |\langle n|U(t)|0\rangle|^2 \ = \ |c_n(t)|^2
\ee
where $U(t)=\exp[-i{\cal H}t]$ is the evolution operator.
The survival probability is $P_0(t)$. 
We can associate with the energy distribution 
a probability density 
\be{0}
\rho_t(\omega) \ \ = \ \ \sum_{n} P_n(t) \ \delta(\omega-\omega_n)
\ee
The energy distribution is characterized by its dispersion 
\be{0}
\Delta E_{\tbox{sprd}}(t) \ = \ \left[ \sum_n\left(E_n-E_0\right)^2P_n(t)\right]^{1/2}
\ee
and by the median $E_{50\%}=E_0$, and also by the $E_{25\%}$ and $E_{75\%}$ percentiles.
The width of the core component is defined as
\be{0}
\Delta E_{\rm core}(t)=E_{75\%}-E_{25\%}
\ee
Our interest below is focused in $P_0(t)$, 
in $\Delta E_{\tbox{core}}(t)$ and in $\Delta E_{\tbox{sprd}}(t)$.
We use the same measures in order to describe 
the LDOS distribution $\rho(\omega)$. 
See Fig.(\ref{fig:LDOScartoon}) for a cartoon 
that illustrates the significance of the 
different measures in the analysis.

The first stage of the analysis is to find the LDOS, 
also known as the strength function. 
In the FM case it can be done analytically using a standard 
Green function method.  In the WM case it is 
possible to generalize the approach by Wigner and followers \cite{wigner,mario}.

The survival amplitude is related 
to the LDOS. Namely, $c_0(t)$ can be written 
as the Fourier transform (FT) of $\rho(\omega)$. 
The derivation of this well known fact 
is simple. Taking the energy reference 
as ${E_0=0}$ we have:   
\be{0}
c_0(t) 
&=& \langle 0|U(t)|0\rangle
\\
&=& \sum_{\nu} \langle 0|\eexp{-i{\cal H}t}
|E_{\nu}\rangle\langle E_{\nu}|0\rangle \nonumber
\\
&=& \sum_{\nu} |\langle E_{\nu}|0\rangle|^2 \ \eexp{-iE_{\nu}t}
\\ 
&=& \sum_{\nu}|\langle E_{\nu}|0\rangle|^2
\ \mbox{FT}[2\pi\delta(\omega-E_{\nu})]
\\
\label{e317}
&=& \mbox{FT}\Big[2\pi\rho(\omega)\Big]
\ee

For obvious reasons there is resemblance 
between the saturation profile $\rho_{\infty}(\omega)$ 
and the LDOS $\rho(\omega)$. In a stochastic (diagonal) 
approximation the former is the auto-convolution
of the latter.
But in order to find the time dependence 
of $\Delta E_{\tbox{core}}(t)$ and $\Delta E_{\tbox{sprd}}(t)$ 
it is not enough to know the LDOS. 
In order to obtain analytical results for the spreading  
we shall use a linear-response strategy 
that can be further refined in the FM case.

\section{Time scales}

There are two frequency cut-offs,
an infrared cutoff ${\omega_{\varrho}=\varrho^{-1}}$
and an ultraviolet cutoff $\omega_c$,
which are the mean level spacing and the 
range of the coupling, respectively.
The associated time scales are the
Heisenberg time $t_{\tbox{H}}$ and
the semi-classical time $t_c$ which are given by
\be{0}
t_{\tbox{H}} &=& 2\pi\varrho \\ 
t_c &=& 2\pi/\omega_c
\ee
We shall see that the {\em continuum} limit ${\varrho^{-1}\rightarrow 0}$ 
is well defined if ${s>0}$. 
If we further assume ${s<2}$ then well defined results 
are obtained also in the {\em universal} limit ${\omega_c\rightarrow\infty}$.  
Thus in the range ${0<s<2}$ we should have 
in the continuum limit a {\em cutoff~free} universal theory, 
that constitutes a generalization of the Wigner decay problem.
Note that for finite values of the cutoffs the actual range is 
more restricted, as explained in Sect.\ref{sect:LDOSF}, 
and depicted in Fig.(\ref{fig:stages}).

We shall see that the decay is characterized by 
what we call {\em generalized Wigner time}:
\be{420}
t_0 &=& 
\left(
\frac{2\pi\epsilon^2}
{\bm{\Gamma}(3{-}s) \sin(s\pi/2)}
\right)^{-{1}/{(2-s)}}  
\ee  
where $\bm{\Gamma}$ is the Gamma function.
The numerical prefactor is explained 
and calculated in Sect.(\ref{sect:P}).
As far as order-of-magnitude estimates 
are concerned a more practical expression is 
\be{421}
t_0 &\approx& 
\left(
\frac{2\pi\epsilon^2}
{(2{-}s)s}
\right)^{-{1}/{(2-s)}}  
\ee  
One observes clearly that $s{\rightarrow}0$ and $s{\rightarrow}2$ 
are limiting cases that require special attention 
due to the increased sensitivity to the infrared and 
to the ultraviolate cutoffs respectively.

\section{The local density of states}

Following the presentation of \cite{wls,wbr} we 
expect the LDOS to have in the most general case 
three regions: 
(i)~The core that consists of the 
levels that are mixed non-perturbatively;  
(ii)~The first order tails; 
(iii)~The higher order tails.
We now explain this terminology  
and the associated phenomenology.

The LDOS is determined by the overlaps $\langle E_{\nu} | 0 \rangle$ 
of the initial state with the perturbed eigenstates.
So naturally the first question that arises 
is whether we can use perturbation theory for the calculation.
First order perturbation theory (FOPT) assumes that we can associate 
perturbed and unperturbed levels (no mixing), 
hence ${\langle E_{\nu{=}0} | n{=}0 \rangle \sim \mathcal{O}(1)}$. 
If this is indeed the case we say that the {\em core} 
consists of one level only, while the tails 
consist of all the $\langle E_{\nu{\ne}0} | n{=}0 \rangle$ overlaps. 
Typically FOPT gives a leading order approximation 
for the tails, while higher orders are essential for those 
levels that are not coupled directly \cite{flamb}. 
In our model the higher order tails reside at ${\omega>\omega_c}$   
out of the range of physical interest, 
because $\omega_c$ is assumed to be very large.

The first order tails are fully determined 
by the spectral function $\tilde{C}(\omega)$.
The self consistent condition for the validity 
of FOPT is ${p_0\ll1}$ where $p_0$ is defined 
as the probability which is carried by the tails:  
\be{0}
p_0 =
\sum_n \left|\frac{V_{n0}}{E_n-E_0}\right|^2 
\ \ = \  \
\int\frac{\tilde{C}(\omega)}{\omega^2} \, \frac{d\omega}{2\pi}
\ee
By substituting the spectral function of Eq.(\ref{e547}) we get 
\be{0}
p_0 = 
\epsilon^2
\left\{\amatrix{
2 \, {\bm \Gamma}(s{-}2) \, {\omega_c}^{s{-}2}  & \mbox{for $s{>}2$}
\cr
(2{-}s)^{-1} \, \varrho^{2{-}s}  & \mbox{for $s{<}2$}
}\right.
\ee
In the continuum limit $p_0$ is infrared divergent if $s<2$,  
and consequently FOPT does not apply even if $\epsilon$ is made very small.
This is the case of our interest, and it should be  
contrasted with the ${s>2}$ case for which FOPT applies 
in the weak coupling limit. In the latter case 
the LDOS can be written schematically as   
\be{14}
\rho(\omega) \ \ = \ \ 
(1{-}p_0)\delta(\omega) 
\ \ + \ \ 
p_0 \, \frac{1}{\omega_c}\tilde{f}\left(\frac{\omega}{\omega_c}\right)
\ee
where the second term stands for $\tilde{C}(\omega)/(2\pi\omega^2)$. 
The natural question that arises is what happens 
to this line shape in the ${s<2}$ case, and in particular 
whether there are remnants of FOPT. It turns out that  
generically the answer is positive \cite{frc}. In spite of 
the mixing of nearby levels, the tails are still given 
by the FOPT expression. Accordingly we can identify 
in the LDOS a {\em core} region ${|\omega|<\gamma_0}$ 
that contains the large overlaps, and FOPT tails 
that dominate the outer ${|\omega|>\gamma_0}$ regions. 
The characteristic frequency can be determined 
either analytically or self-consistently and  
consequently the generalized Wigner time is identified 
as ${t_0=1/\gamma_0}$. This phenomenological picture 
will be formulated in a more precise way in the subsequent subsections.

We are going to analyze the LDOS for ${s<2}$, 
where we have to go beyond FOPT, but still can use 
the core-tail phenomenology.   
The non-perturbative core within ${|\omega|<\gamma_0}$ 
has non-universal structure that depends 
on the details of the model.
There are various strategies to deduce 
the structure of the core. In particular:   
the Green function method that we 
are going to use in the FM case;   
the RMT calculation of the moments as in the 
pioneering work of Wigner \cite{wigner,mario};  
and the semiclassical reasoning if the model 
has a classical limit \cite{wls,lds,brm,HKG}.

A few words about the Lamb shift are in order.
FOPT allows to calculate the shift of $E_0$  
due to the repulsion by the other levels: 
\be{0}
\Delta(0) &=& 
\sum_n\frac{\left|V_{n,0}\right|^2}{E_0-E_n}
=  -\int_{-\omega_c}^{+\omega_c}\frac{\tilde{C}(\omega')}
{\omega'}\frac{d\omega'}{2\pi}
\ee
If $s>0$ the lamb shift is {\em not} infrared 
divergent, provided we keep away from the 
energy floor, and we get $\Delta(0)=0$ due to the 
symmetry of $\tilde{C}(\omega)$.
This of course does not mean that Lamb-shift physics 
is irrelevant. 
In the analysis of the FM model we shall define 
a spectral function $\Delta(\omega)$ that plays a 
major role in the analysis.

\subsection{The LDOS - FM}
\label{sect:LDOSF}

In the FM case it is possible to derive an exact expression 
for the LDOS either via the Green function formalism (App.A)
or from a straightforward elementary calculation (App.B).
The final result is 
\be{20}
\rho(\omega)  = \frac{1}{\pi}\, \frac{\Gamma(\omega)/2}
{(\omega-\Delta(\omega))^2+(\Gamma(\omega)/2)^2}
\ee
where
\be{0}
\Gamma(\omega) &=& 
\sum_n \left|V_{n,0}\right|^2 2\pi\delta\left(E-E_n\right)
\ = \ \tilde{C}(\omega) \\
\label{e18}
\Delta(\omega) &=& 
\sum_n\frac{\left|V_{n,0}\right|^2}{E-E_n}
\ = \ \int_{-\infty}^{+\infty}\frac{\tilde{C}(\omega')}
{\omega-\omega'}\frac{d\omega'}{2\pi}
\ee

\begin{figure}[h!]
\begin{center}
\includegraphics[width=0.99\hsize,clip]{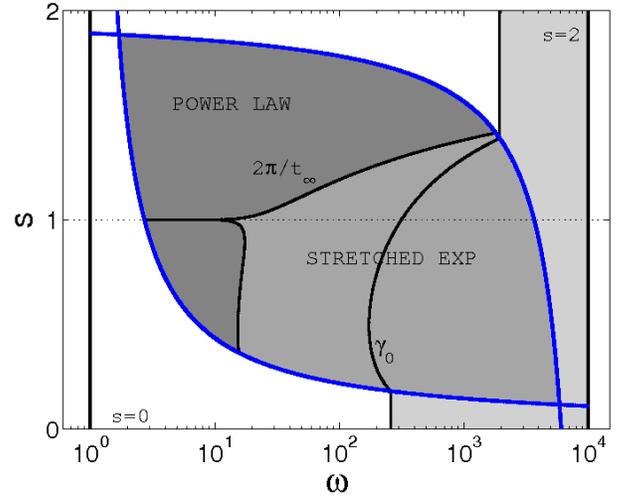}
\end{center}
\caption{(color online) 
The different $\omega$ regions of the LDOS are plotted
for $0{\leq}s{\leq}2$, relating to the (richer) FM case. 
The universal $\omega$ regions where the LDOS is cutoff-free,  
are shaded. The implied time-domain decay is indicated
(note that upon the identification ${t\sim1/\omega}$ 
one can regard the horizontal axis as time stretching from right to left). 
The lower and upper cut-offs $\omega_{\varrho}$ and $\omega_c$, are indicated 
by vertical solid thick lines. The border of the different $\Delta(\omega)$ 
expressions written in Eq.(\ref{e34}) are plotted 
in curved solid (blue) lines.
The curves of $\gamma_0$ and $2\pi/t_\infty$ are plotted as well.
The $\gamma_0$ curve was slightly modified for the purpose
of presentation. The parameters used in this plot 
are $\omega_{\varrho}{=}1$ and $\omega_c{=}10^4$ and $\epsilon{=}2$.
}
\label{fig:stages}
\end{figure}

In the Ohmic case ($s{=}1$) we have ${\Delta(\omega)= 0}$, 
and ${\Gamma(\omega) = \mbox{const}}$, and consequently 
the LDOS is the familiar Lorentzian.
In order to calculate $\Delta(\omega)$ for ${s\ne1}$ we 
exploit the fact that $\tilde{C}(\omega)$ 
is an even function, while $\Delta(\omega)$ comes out odd.
Consequently we can write Eq.(\ref{e18}) as
\be{4}
\Delta(\omega) &=& \epsilon^2 \ \frac{\omega}{\pi}\int_{\omega_{\varrho}}^{\omega_c}
\frac{(\omega')^{s{-}1} d\omega'  }{\omega^2-\omega'^2}
\\ \nonumber
&=& -\epsilon^2 |\omega|^{s-1}\mbox{sgn}(\omega)
\int_{\mbox{ln}|\omega_{\varrho}/\omega|}^{\mbox{ln}|\omega_c/\omega|}
\frac{\eexp{(s-1)x}}{\sinh(x)} \, \mbox{d}x
\ee
where $\omega_{\varrho}=\varrho^{-1}$ is the level spacing, 
and we used the substitution $\omega'=\omega\eexp{x}$. 
For $0{<}s{<}2$ it is possible to take the 
limits $\omega_{\varrho}{\rightarrow}0$ 
and $\omega_c{\rightarrow}\infty$.
Using the integral
\be{0}
\int_{-\infty}^{+\infty}
\frac{\sinh((s-1)x)}{\sinh(x)} 
\, \mbox{d}x 
\ = \ 
{-}\pi\cot(s\pi/2)
\ee
we get the cutoff-free result 
\be{6}
\Delta(\omega) = 
\epsilon^2\pi\cot\left(s{\pi}/{2}\right)|\omega|^{s-1}\mbox{sgn}(\omega) 
\ee
In contrast to that, the marginal cases $s{=}0$ and $s{=}2$ 
are lower and upper cut-off dependent respectively:
\be{0}
\Delta(\omega) =& 
-\epsilon^2 \, \omega \, \ln\left|1-\left({\omega_c}/{\omega}\right)^2\right|,
& \ \ \ \ \ \ s{=}2 
\\
\label{e7}
\Delta(\omega) =& 
\epsilon^2 \, \frac1\omega \, \ln\left|\left({\omega}/{\omega_{\varrho}}\right)^2-1\right|,
& \ \ \ \ \ \ s{=}0 
\ee
The crossover from the ``$0{<}s{<}1$'' to the  ``$s{=}0$'' and the ``$s{=}2$''
expressions is not sharp. Looking carefully at the integral we see that the 
condition for a cutoff independent result is (see Fig.\ref{fig:stages}):
\be{34}
\omega_{\varrho} \, \eexp{1/s}
\ \ \ll \ \ 
\omega 
\ \ \ll \ \
\omega_c \, \eexp{-1/(2{-}s)}
\ee

\begin{figure}[h!]
\begin{center}
\includegraphics[width=0.85\hsize,clip]{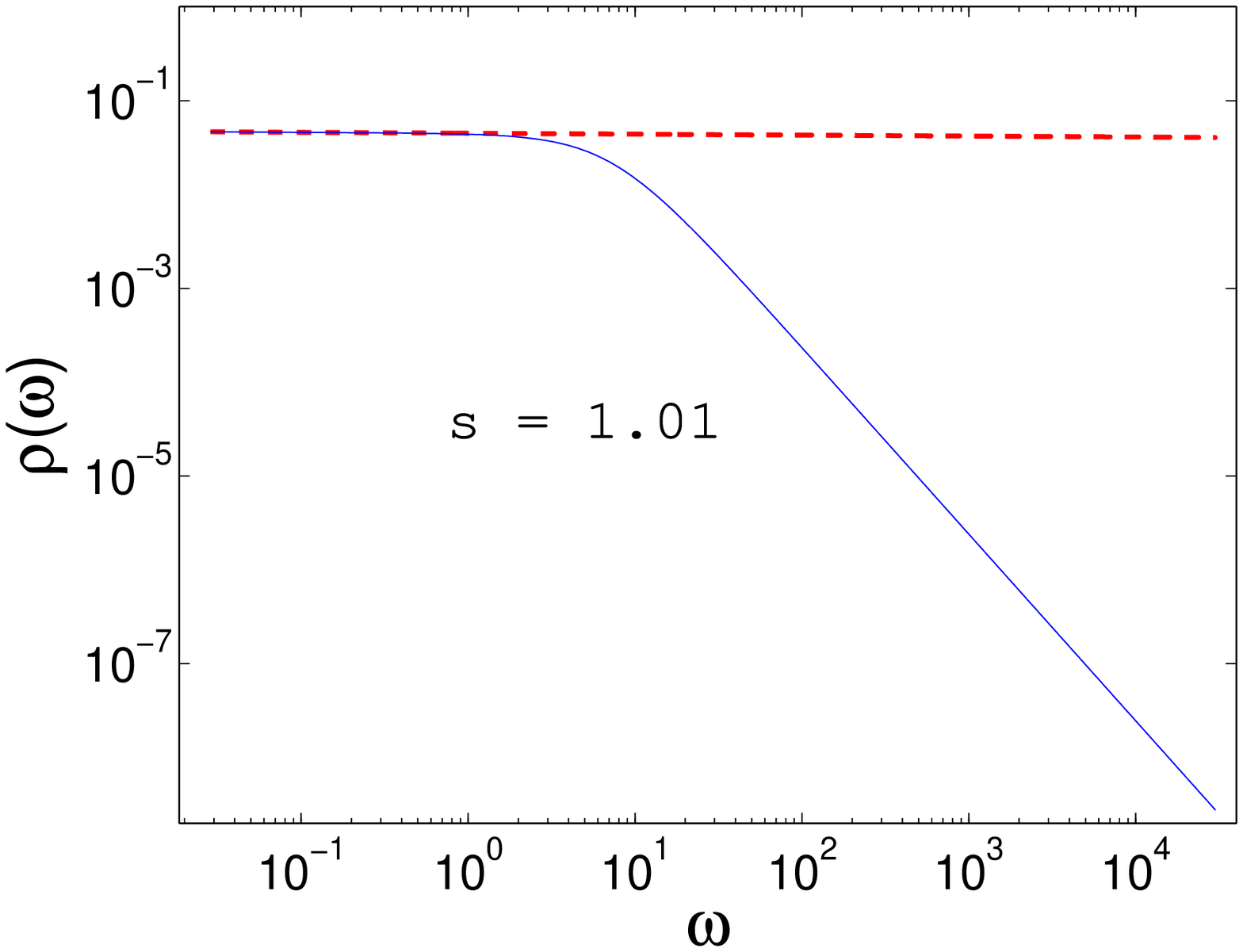}\\
\includegraphics[width=0.85\hsize,clip]{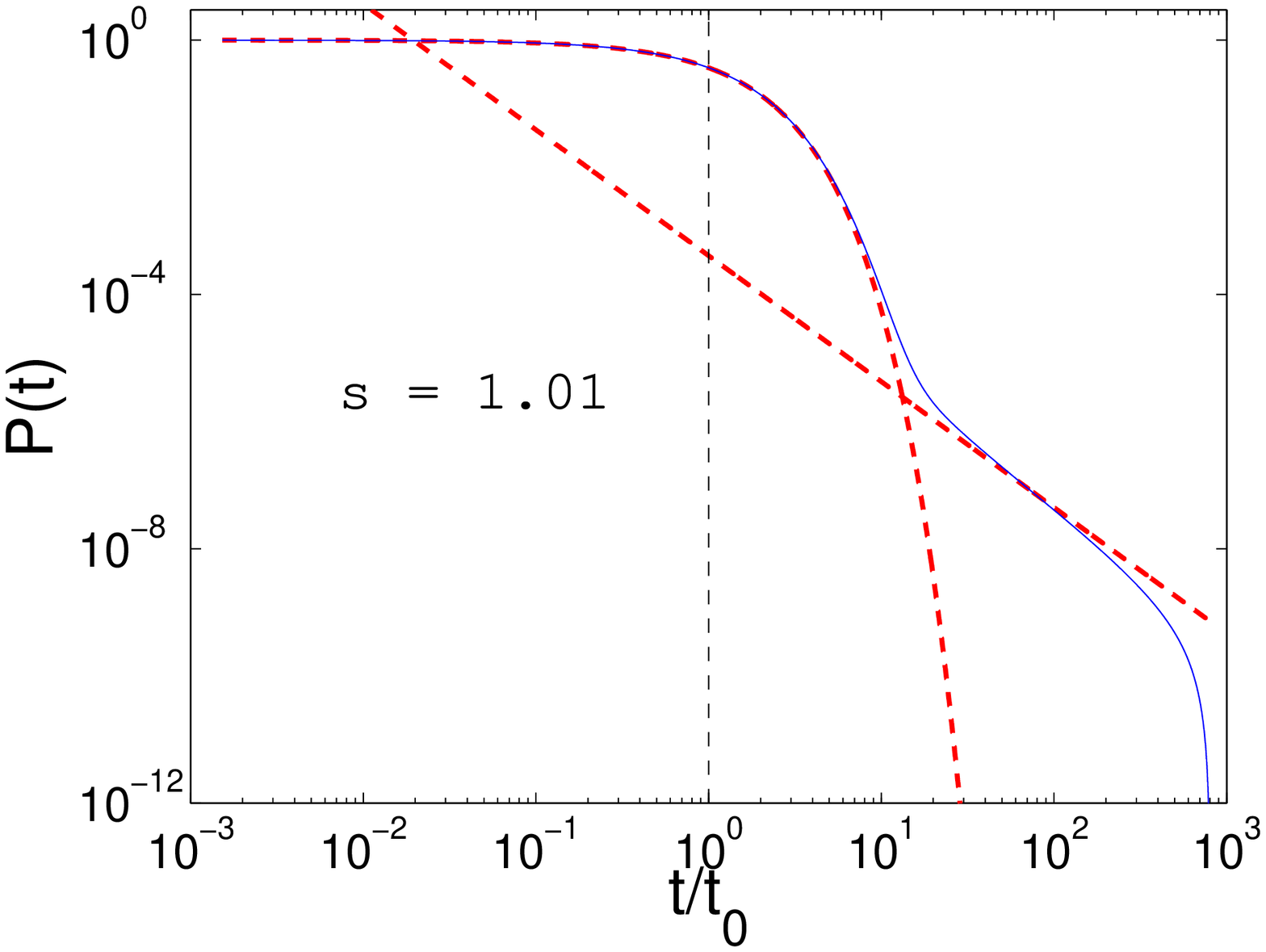}\\
\includegraphics[width=0.9\hsize,clip]{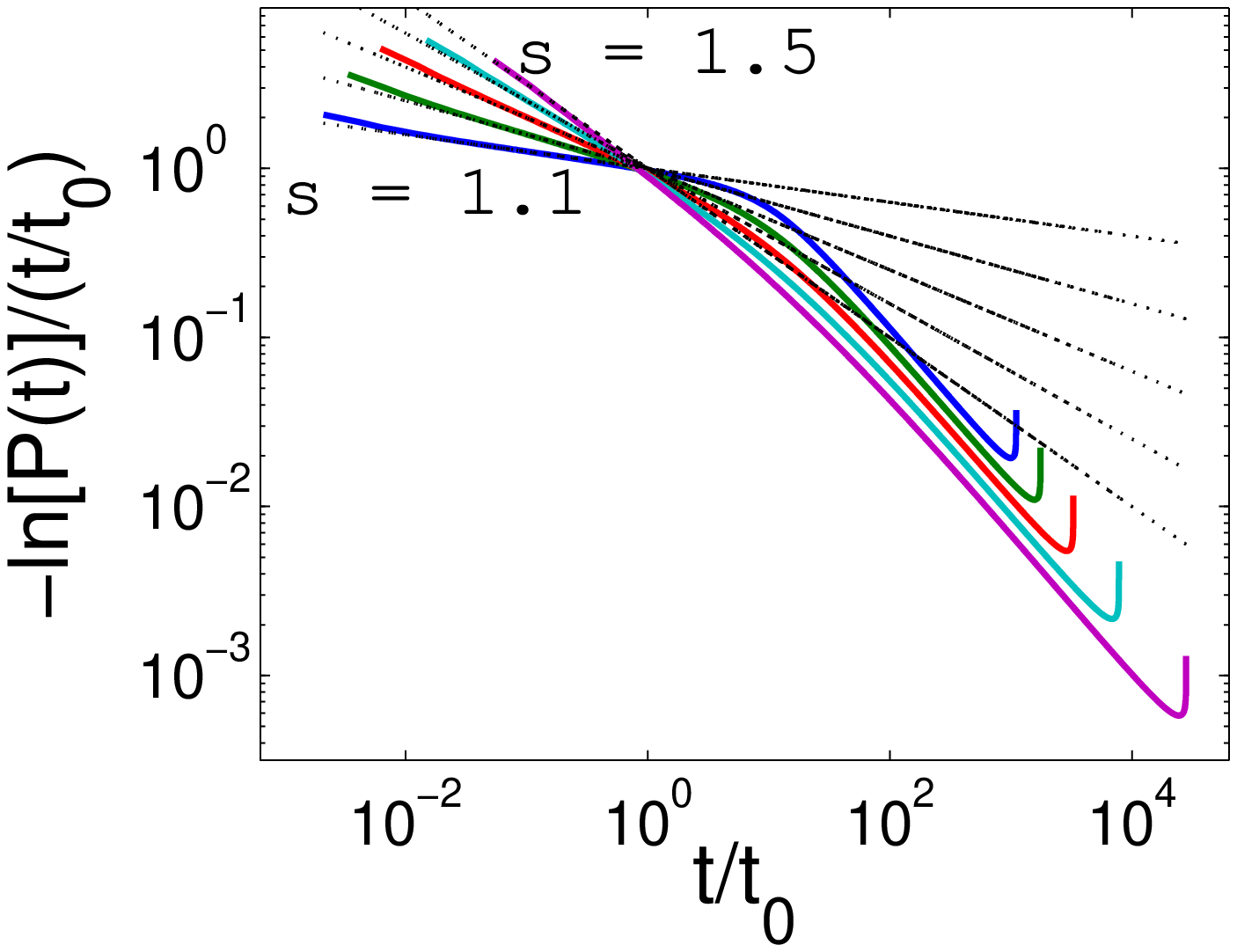}
\end{center}
\caption{(color online) 
\textit{Upper panel:} The FM analytical LDOS for $s{=}1.01$.
The core power-law (Eq.(\ref{e27})) is indicated by a dashed red line.
\textit{Middle panel:} The survival probability $P_0(t)$
computed using numerical FT of the LDOS.
The exponential-like decay (based on Eq.(\ref{e39})) 
and the power-law decay (based on Eq.(\ref{e40}))
are plotted in dashed red lines.
The vertical dashed black line emphasis the fact
that the cross-over time $t_{\infty}$ is different from $t_0$.
\textit{Lower panel:} The same for various 
values of ${s\in[1.1,1.5]}$.
For short times there is a good agreement with an
exponnential-like decay (based on Eq.(\ref{e39})),
which is plotted in dotted black lines.
}
\label{fig:FT}
\end{figure}

\begin{figure}[h!]
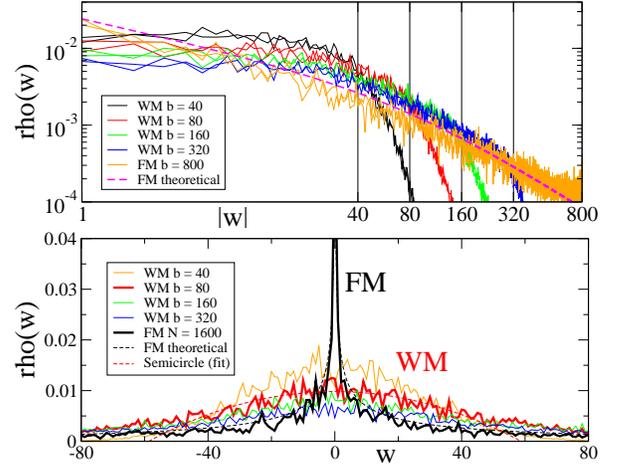

\begin{center}
\includegraphics[width=0.9\hsize,clip]{kbs_ldos}
\includegraphics[width=0.9\hsize,clip]{kbs_ldos_linearScale}
\end{center}
\caption{(color online) 
The LDOS for the FM and for the WM via 
direct diagonalization of $1600\times1600$ matrices 
with ${s=1.5}$ and ${\epsilon=1.44}$.
The units are such that $\varrho{=}1$,  
and hence the bandwidth is ${\omega_c=b}$.    
In the FM case formally $b{=}N/2$.
{\it Upper panel:} The log-log scale 
emphasizes the emergent universality of the tails 
as the cutoff~$\omega_c$ is taken to infinity. 
{\it Lower panel:} The log-linear scale emphasizes 
the difference in the non-universal core component:
FM has a singular core, while WM has a smooth semicircle-like core. 
}
\label{fig:ldos}
\end{figure}

Having obtained $\Delta(\omega)$ we can substitute it into
Eq.(\ref{e20}) and get the LDOS.
Both $\Gamma$ and $\Delta$ are $\propto\omega^{s-1}$, 
with $s$~dependent prefactors.  
One can define a characteristic crossover 
frequency $\gamma_0$ above which 
the $\omega^2$ term in the denominator 
if Eq.(\ref{e20}) dominates. In the universal 
regime this leads to 
\be{0}
\gamma_0 \approx \left(\frac{\epsilon^2}{|\sin(s\pi/2)|}\right)^{1/(2-s)}
\ee
while in the limiting cases it attains the values 
\be{536}
\gamma_0 \approx & \omega_c\eexp{-\frac{1}{2\epsilon^{2}}} & \ \ \mbox{as $s{\rightarrow}2$}
\\
\gamma_0 \approx & \epsilon\ln(\varrho\epsilon) & \ \ \mbox{as $s{\rightarrow}0$}
\ee
Depending whether ${\omega\ll\gamma_0}$ 
or ${\omega\gg\gamma_0}$ one obtains 
different approximations for the LDOS 
that can be packed together 
in the following writing style 
\be{0}
\rho(\omega)  = \frac{1}{\pi}\,
\left\{
\frac{\Gamma/2}{\Delta^2+(\Gamma/2)^2}, 
\ \ 
\frac{\Gamma/2}{\omega^2}\right\}
\ee
Consequently we get for $0{<}s{<}2$ 
\be{27}
\rho(\omega) = 
\left\{
\left(\frac{\sin(s\pi/2)}{\pi\epsilon}\right)^2\frac{1}{|\omega|^{s-1}}, 
\ \ 
\frac{\epsilon^{2}}{|\omega|^{3-s}}
\right\}
\ee
This validity regions of the expression in the curly brackets 
are illustrated in Fig.(\ref{fig:stages}). After FT they 
are reflected in time domain as implied by the analysis 
in Fig.(\ref{fig:FT}), which we further discuss later.
The numerically determined LDOS in the FM case
is displayed in Fig.(\ref{fig:ldos}), 
and contrasted with the WM case.

Eq.(\ref{e27}) holds provided $\gamma_0$ 
is in the range that has been specified in Eq.(\ref{e34}).
This constitutes a condition on $\epsilon$ 
that can be regarded as a generalized version 
of the FGR condition. 
For marginal values of $s$ it becomes more difficult to 
satisfy the generalized FGR condition, 
and eventually the range of the universal behavior 
as defined in Eq.(\ref{e34}) shrinks to zero (see Fig.(\ref{fig:stages})). 
In particular for $s{\rightarrow}2$,  
once the generalized FGR condition breaks down, we get 
\be{540}
\rho(\omega) = 
\left\{
\frac{1}{4\epsilon^2} \, \frac{1}{|\omega|\ln^2\left|\omega_c/\omega\right|}, 
\ \ 
\frac{\epsilon^{2}}{|\omega|}
\right\}
\ee
In this expression the $1/\omega$ tails prevail 
for ${\omega > \gamma_0}$, 
analogous to Eq.(\ref{e27}), 
but here $\gamma_0$ is $\omega_c$ dependent.

\subsection{The LDOS - WM}

As discussed in the beginning of this section, 
for ${s>2}$ we can calculate the LDOS using FOPT.  
In contrast, for ${s<2}$, FOPT breaks down 
for any finite coupling, because in the continuum 
limit $p_0$ is infrared divergent.
Still we can use the FOPT result 
down to a frequency $\gamma_{o}$ 
which we estimate below in a self consistent manner. 
For lower frequencies the LDOS is dominated by 
a non-perturbative core which is model specific
and in that sense, non-universal. 
For the WM we write schematically  
\be{0}
\rho(\omega) \approx
\left\{
{\mbox{Semi-circle}}, 
\ \ 
{\frac{\epsilon^2}{|\omega|^{3-s}}}
\right\}
\ee
This functional form is illustrated 
in Fig.(\ref{fig:LDOScartoon}), 
and tested numerically in Fig.(\ref{fig:ldos}). 
The reasoning behind this expression 
is further detailed below.

The crossover frequency $\gamma_{o}$ is almost 
the same as $\gamma_{0}$ which we define 
in the time domain analysis of the next section 
[the distinction between the two is clarified after Eq.(\ref{e701})].    
We define the crossover frequency $\gamma_{o}$ by the equation
\be{542}
\int_{|\omega|>{\gamma_{o}}} \frac{\tilde{C}(\omega)}{\omega^2}
\, \frac{d\omega}{2\pi}\sim50\%
\ee
which gives
\be{543}
\gamma_{o} \ \approx \ \left(\frac{\epsilon^2}{2{-}s}\right)^{1/(2{-}s)}
\ee
The FOPT tails are dominated by ${\cal H}_0$, 
while $V$ is regarded as a perturbation. 
In contrast to that the core is fully dominated by~$V$. 
Following Ref.\cite{mario} we can argued 
that the core of the LDOS is a semicircle 
of width $\Delta E_{\tbox{sc}}$, which in our case 
is given by the expression 
\be{0}
\Delta E_{\tbox{sc}} = \left[\int_0^{{\gamma_{o}}} \tilde{C}(\omega)
\, \frac{d\omega}{2\pi} \right]^{1/2}
\ee
In this expression we use the effective bandwidth $\gamma_{o}$ 
rather than the real bandwidth $\omega_c$, 
as implied by our core-tail hypothesis.  
Performing the integration we get
\be{0}
\Delta E_{\tbox{sc}} \ \sim \ \gamma_{o}
\ee
which implies that the calculation is indeed self-consistent, 
i.e. there is a well defined frequency $\gamma_{o}$ 
at which the core-tail crossover takes place.

\section{The decay of the survival probability}

We turn now to the study of wavepacket dynamics, with the objective 
to contrast the results of the FM with those of the generalized WM. 
We assume that the initial preparation is delta-like 
in energy space. The temporal behavior of the wavepacket,
is characterized by three quantities: 
(a)~the survival probability is directly related to the LDOS analysis; 
(b)~the core width reflects the non-universal component of the wavepacket; 
(c)~the energy spreading reflects the tails of the wavepacket.

In the present section we study the survival amplitude, 
which is obtained via FT of the LDOS, 
namely ${c_0(t)=\mbox{FT}[2\pi\rho(\omega)]}$.  
For $s{>}2$ the FT of the FOPT estimate Eq.(\ref{e14})
has a simple crossover at $t_c$ to saturation
\be{0}
c_0(t) &=& 
(1{-}p_0)+p_0 \, f(t/t_c) \nonumber \\ 
&\sim& 1 - p_0\, \left[1- \frac{1}{(t/t_c)^{s{-}2}} \right] 
\ee
For $s{<}2$ the core is not a discrete delta function 
but rather has a width $\gamma_{o}$ which implies 
a progressive decay  on time scale ${1/\gamma_{o}}$.
Our focus below is on the regime $0{<}s{<}2$, and we 
pay some extra attention to the limiting cases $s{\rightarrow0}$ 
and $s{\rightarrow2}$.

\subsection{The survival probability - WM}
\label{sect:P}

In the WM case the LDOS has power-law tails $1/|\omega|^{1{+}\alpha}$
with ${\alpha=2{-}s}$, and a smooth semicircle core. 
Therefore its FT is exponential-like for all times. 
The similarity to the $\alpha$-stable Levy distribution 
suggests a stretched exponential decay which is confirmed by 
our simulations. 
There is an optional argument that supports a stretched 
exponential behaviour. 
The stretched exponential can be regarded as
the solution of the following phenomenological 
rate equation for the survival probability
\be{0}
\dot{\mathcal{P}}(t) = -\Gamma(\const/t) \ P_0(t)
\ee
This phenomenological equation
involves a time dependent effective decay rate
into the quasi continuum. 
It is estimated as $\Gamma(\omega)$ 
with ${\omega \sim 1/t}$ which reflects
the energy uncertainty at time~$t$.  
This phenomenological approach does 
not provide a definite result for the 
numerical $\const$. A reasonable 
procedure is to determine this constant  
from the short time behavior, 
which is determined by the tails. 
We recall the following FT relation:
\be{648}
\int_{-\infty}^{\infty} \!\!\!
|\omega|^{\beta{-}1} {\cos(\omega t)}
\, \frac{\mbox{d}\omega}{2\pi}
=   
\left[\frac{1}{\pi} \, \Gamma(\beta) \,
\cos\left(\beta\frac{\pi}{2}\right)\right]
\frac{1}{|t|^{\beta}} 
\ \ \ \ \ 
\ee
where $\beta>0$. Switching the roles 
of $t$ and $\omega$, the inverse FT implies 
that power law tails $1/|\omega|^{1{+}\alpha}$
are related to discontinuity or 
singularity $C|t|^{\alpha}$ at time domain, 
where ${C=-[2\Gamma(1{+}\alpha)\sin(\alpha\pi/2)]^{-1}}$.
This leads to the result 
\be{39}
c_0(t) \ \ = \ \ \exp\left[-\frac{1}{2}\left(\frac{t}{t_0}\right)^{2{-}s}\right] 
\ee
where $t_0$ is defined as in Eq.(\ref{e420}). 
Going back to the phenomenological rate equation 
it implies that the numerical constant 
is $[(2{-}s)/(\bm{\Gamma}(3{-}s) \sin(s\pi/2))]^{1/(s{-}1)}$.

\subsection{The survival probability - FM}
\label{sect:PF}

In the FM case the core is singular and therefore 
a power-law takes over for $t>t_{\infty}$, 
where the crossover time $t_{\infty}$ is determined below.
In Fig.(\ref{fig:FT}) we present numerical FTs that
support this statement. Note that in the $s{=}2$ limit 
the power law becomes a logarithmic decay. 

\begin{figure}[h!]
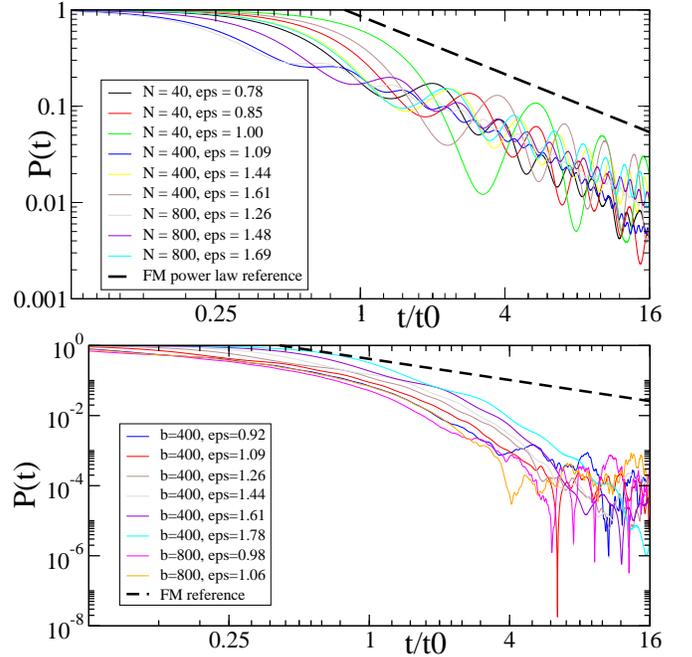

\begin{center}
\includegraphics[width=1\hsize,clip]{kbs_Pt_FM}\\
\includegraphics[width=1\hsize,clip]{kbq_Pt_WM}
\end{center}
\caption{(color online) 
The survival probability $P_0(t)$ for the FM (\textit{top})
and for the WM (\textit{bottom}), as a function of ${t/t_0}$, 
in log-log scale, for various values of $\varepsilon$. 
The predicted FM power-law is presented as a dashed line.  
For all curves ${\varrho=1}$ and ${s=1.5}$.
}
\label{fig:P}
\end{figure}

\begin{figure}[h!]
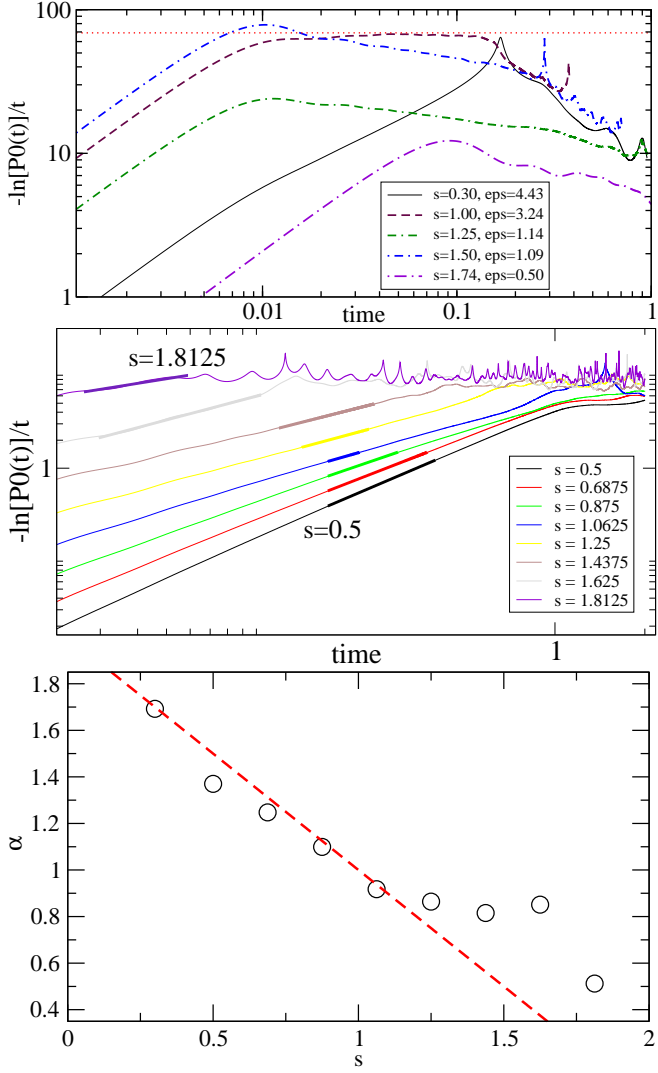

\begin{center}
\includegraphics[width=1\hsize,clip]{kbq_Pt_WM_Inset}
\includegraphics[width=1\hsize,clip]{kbq_stretch}
\includegraphics[width=1\hsize,clip]{extracted}
\end{center}
\caption{(color online) 
\textit{Upper panel:} 
A re-plot of Fig.(\ref{fig:P}) using $Y=-\ln[P_0(t)]/t$ and $X=t$
in a log-log scale, for representative runs, 
showing that the decay in the WM case is described 
by a stretched exponential.
The red bold dashed line has zero slope, 
corresponding to simple exponential decay for $s{=}1$.
\textit{Middle panel:} 
Additional curves for various values of $s$ are plotted. 
The highlighted (thickened) segments demonstrate exponential-like decay. 
\textit{Lower panel:}
The power $\alpha$ is extracted by fitting to the form $P_0(t)=\exp{[-t^\alpha]}$
of the highlighted segments of the middle panel. 
The expected power $\alpha{=}2{-}s$ is indicated by a dashed red line.
}
\label{fig:stretch}
\end{figure}

In order to obtain the explicit expression for 
the power law decay we use the FT relation of Eq.(\ref{e648}).
Due to the discontinuity or singularity 
at ${\omega=0}$ we get for large times
\be{40}
c_0(t) &=& \frac{\sin((s{-}1)\pi)\sin(s\pi/2)\Gamma(2{-}s)}
{\pi^2\epsilon^2 \ t^{2{-}s}}\nonumber\\
&=&  \frac{2\sin((s{-}1)\pi)}{(2{-}s)\pi} 
\ \left(\frac{t_0}{t}\right)^{2{-}s}
\ee
where in the last equation we expressed $c_0(t)$ by $t_0$.
Comparing the exponential-like and the power law
we find an expression for the crossover time 
\be{0}
t_{\infty} \ = \ \left|
2\ln\left(\frac{2\sin(|s{-}1|\pi)}{(2{-}s)\pi}\right)
\right|^{{1}/{(2{-}s)}}t_0
\ee
This expression diverges in the limit ${s=1}$, 
implying that only the exponential survives. 
The numerical simulations presented in Figs.(\ref{fig:P})
and (\ref{fig:stretch}) support our findings.

The marginal case $s{=}2$ requires special 
treatment. In this case the strength of the 
interaction ($\epsilon$) is a dimensionless parameter.  
The ${\propto 1/\omega}$  tail of the LDOS in Eq.(\ref{e540}) 
is $\omega_c$ independent, but with the  
$\omega_c$ dependent lower cutoff of Eq.(\ref{e536}).
For the purpose of FT we approximate the LDOS as 
\be{0}
\rho(\omega)\approx 
\frac{\epsilon^2}{\sqrt{\omega^2+(1/t_0)^2}}
\ee
and from the integral representation of the 
modified Bessel function of the second kind,
using the approximation ${K_0(\tau) \approx \const-\ln(\tau)}$ 
it follows that 
\be{0}
c_0(t) \ \ \approx \ \ \frac{\epsilon^2}{\pi}\log\left(\frac{t_0}{t}\right)
\ \ \ \ \mbox{for} \ \ t_c \ll t \ll t_0
\ee
In order to clarify how the $s{=}2$ case 
is related to the general $s$ expression,  
it is useful to note that for very small $\beta$ 
one has ${t^{\beta} \sim  \beta \ln(t)}$. 
It is also important to realize that for $t<t_0$ 
both the $s{=}1$ exponential and the $s{=}2$ logarithm   
are consistent with the short time $1-(t/t_0)^{2{-}s}$ 
expression that holds for general~$s$. 
In contrast to that the long time decay is 
sensitive to the small frequency features of the LDOS.

\section{The evolution of the core width}

\begin{figure}[h!]
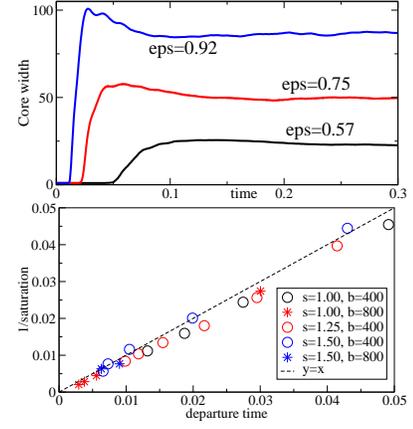

\begin{center}
\includegraphics[width=0.6\hsize,clip]{WMn50dep2} \\
\includegraphics[width=0.6\hsize,clip]{WMn50dep1}
\end{center}
\caption{(color online) 
{\em Upper panel:} Examples for the time evolution of $\Delta E_{\tbox{core}}$ 
for $s{=}1.5$ and $b{=}800$ in the WM case. 
{\em Lower panel:} The extracted departure time versus the extracted
inverse saturation value. This scatter diagram demonstrates the validity 
of one parameter scaling.
}
\label{fig:WMn50dep}
\end{figure}

In this section we study the time evolution 
of the core as reflected in the width of the energy distribution. 
The width $\Delta E_{\tbox{core}}(t)$ is intimately related to $P_0(t)$. 
It starts to rise at $t\sim t_0$ when $P_0(t)=50\%$, 
and its saturation value is determined by the core width of the LDOS $\sim\gamma_0$. 
Thus $\Delta E_{\tbox{core}}$ should exhibit 
one parameter scaling with respect to~$t_0$.
This has been verified numerically, and is shown in Fig.(\ref{fig:WMn50dep}).
A more elaborated analysis follows below.

For $t\ll t_0$ the initial state carries most of the probability, 
and therefore we can use FOPT to obtain in analogy with Eq.(\ref{e14}):
\be{0}
\rho_t(\omega) \ \ = \ \ 
\delta(\omega) 
\ \ + \ \ 
\frac{1}{2\pi}
\frac{\tilde{C}(\omega)}{\omega^2} [2\sin(\omega t/2)]^2
\ee
with normalization measure $d\omega$.
As long as this expression holds 
we say that the core component contains one level only. 
It is clear that after a longer time $\rho_t(\omega)$ becomes 
similar to the LDOS $\rho(\omega)$.

Accordingly it is natural to ask what is 
the time evolution of the core-tail structure.  
In particular one would like to estimate  
the time dependence of the border~$\gamma(t)$ 
between the core and the tail.
Using the same reasoning as in Eq.(\ref{e542}) 
we get for~$\gamma(t)$ the equation 
\be{701}
\epsilon^2 \times \left[ \frac{1}{s}(t^{2-s}-\gamma^st^2) +  \frac{1}{2{-}s}(t^{2{-}s})\right] \ \sim \ 50\%  
\ee
From this equation it follows that there is 
a sharp crossover from  $\gamma(t)=\varrho^{-1}$ 
to $\gamma(t)\sim \gamma_{o}$ during the 
time interval ${[t_0,t_o]}$ where $t_0$ is given  
by Eq.(\ref{e421}) and ${t_o=1/\gamma_o}$ is 
given by Eq.(\ref{e543}).
In Fig.(\ref{fig:WMn50dep}) we present the results of the numerical analysis.
Our data, indicate that the expected one-parameter scaling 
is obeyed. We have verified that the deviations from the 
expected $\epsilon$ dependence (for large~$\epsilon$) 
are an artifact due to having finite (rather then infinite) bandwidth.

\section{The time dependent spreading}

In this section we analyze a very different characteristic of the
evolving wavepacket, which is the spreading $\Delta E_{\tbox{sprd}}(t)$.
Unlike the width which is determined by the core, 
the spreading is determined by the tails of the distribution.
Irrespective of whether we deal with FM or with WM the tails 
are determined by FOPT and therefore the naive expectation 
is to have a $t_c$ dependent rather than $t_0$ dependent evolution.  
We shall see that this is roughly but not quite correct.

\begin{figure}[h!]
\begin{center}
\includegraphics[width=0.8\hsize,clip]{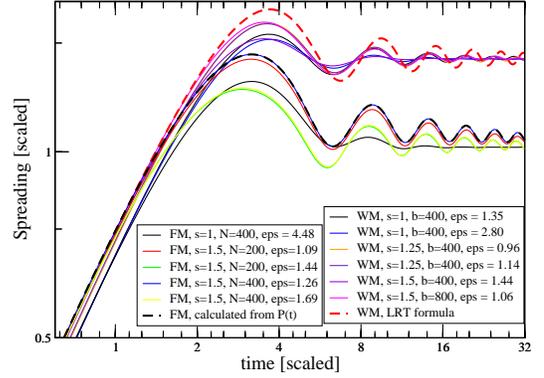}
\end{center}
\caption{(color online) 
Scaled spread $\Delta E_{\tbox{sprd}}/(\omega_c^s\epsilon^2/s)^{1/2}$
versus scaled time ${\omega_c t}$ for the FM and the WM. 
The theoretical predictions Eq.(\ref{e14}) and Eq.(\ref{e18}) 
are plotted for comparison.
}
\label{fig:spreading} 
\end{figure}

For the analysis one can use the traditional strategy of Refs.\cite{wbr,brm,ref20}, 
leading to the following spreading formula (see App.C): 
\be{46}
\left[\Delta E_{\tbox{sprd}}(t)\right]^2 = C(0,0) + C(t,t)- C(0,t) -C(t,0)
\ee
where 
\be{0}
C(t', t'') \ \  = \ \ \langle V(t')V(t'')\rangle
\ee
This expression is formally exact. But in order to get a practical 
expression the LRT approximation assumes that it is possible 
to calculate the correlation function with the unperturbed Hamiltonian.
Accordingly ${C(t', t'') \approx C(t'-t'')}$ where $C(t)$ 
is the inverse FT of $\tilde{C}(\omega)$, leading to  
\be{47}
\Delta E_{\tbox{sprd}}(t) = \left[ 2\Big(C(0)-C(t)\Big)\right]^{1/2}
\ee
This expression implies that 
\be{477}
\Delta E_{\tbox{sprd}}(\infty) \ = \ \left[ 2 \times \frac{1}{s}\omega_c^s \ \epsilon^2\right]^{1/2}
\ee
This saturation values, which diverges in the ${\omega_c\rightarrow\infty}$ limit, 
is attained after a short transient time ${t_c \rightarrow 0}$, 
while the generalized Wigner time~$t_0$ is not reflected.
The above expressions agrees with the numerical calculations,
as shown in Fig.(\ref{fig:spreading}).

The LRT approximation ${C(t', t'') \approx C(t'-t'')}$ makes sense if 
the correlation function is stationary. This is not the case with FM.
By inspection a more appropriate approximation in the latter case 
should take into account the decay of the initial state
and hence it should incorporate the survival amplitude $c_0(t)$ as follows:
\be{0}
C(t', t'') \ \  \approx \ \ c_0(t') \, c_0(t'') \, C(t'-t'')
\ee
Within the framework of this approximation we get 
\be{55}
\Delta E_{\tbox{sprd}}(t) \approx \Big[(1{+}P_0(t))C(0) - 2c_0(t)C(t)\Big]^{1/2}
\ee
where $P_0(t)=|c_0(t)|^2$. This leads to a  saturation value smaller 
by factor $\sqrt{2}$ compared with the WM case, 
reflecting the non-stationary decay of the fluctuations as a function of time.
More interestingly Eq.(\ref{e55}) contains a cutoff independent
term that reflects the universal time scale $t_0$.

\begin{figure}[h!]
\begin{center}
\includegraphics[width=0.8\hsize,clip]{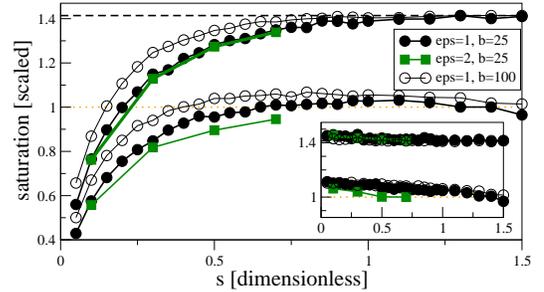}
\end{center}
\caption{(color online) 
The scaled saturation value of $\Delta E_{\tbox{sprd}}$
for the FM (lower curves approach~$1$) 
and for the WM (upper curves approach~$\sqrt{2}$).
The saturation value falls down as~$s$ becomes smaller 
due to the finite level spacing. To establish 
the latter statement we re-plot the same data 
but re-scale the spread 
as $\Delta E_{\tbox{sprd}}/[(\omega_c^s\epsilon^2/s)-(\omega_{\varrho}^s\epsilon^2/s) ]^{1/2}$.
}
\label{fig:sat} 
\end{figure}

In fact in the FM case it is possible to obtain an exact result. 
First we notice that by definition $c_0(t)=\langle\eexp{-i{\cal H}t}\rangle$, 
where the expectation value here and below is with 
respect to the initial $|0\rangle$ state. By differentiating we get 
either ${\dot{c}_0}(t)= -i\langle{\cal H} \, \eexp{-i{\cal H}t}\rangle$, 
or ${\dot{c}_0}(t)=-i\langle\eexp{-i{\cal H}t} \, {\cal H}\rangle$. 
Setting ${\cal H}={\cal H}_0+V$ and using the convention $E_0{=}0$ for 
the $|0\rangle$ eigenstate of ${\cal H}_0$ we find 
\be{0}
{\dot{c}_0}(t) &=& -i\langle V \, \eexp{-i{\cal H}t}\rangle
\ \ = \ \ -i\langle\eexp{-i{\cal H}t} \, V\rangle
\ee
Similarly we can handle the second derivative leading to 
\be{0}
{\ddot{c}_0}(t) &=& -\langle{\cal H} \, \eexp{-i{\cal H}t} \, {\cal H}\rangle
\ \ = \ \ -\langle V \, \eexp{-i{\cal H}t} \, V\rangle
\ee
We now realize that the correlations in Eq.(\ref{e46}) 
can be expressed using $C(t)$ and $c_0(t)$. 
Trivially we have 
\be{0}
\langle V(0)V(0)\rangle &=& \langle V^2\rangle = C(0) 
\ee
In the FM model the only nonzero elements of $V$ are $V_{0,n}$ and $V_{n,0}$ 
with $n\ne0$. Consequently we can factorize also the 
other correlations as follows:
\be{0}
\langle V(t)V(t)\rangle &=&
\langle \eexp{i{\cal H}t}\rangle \langle V^2\rangle  \langle \eexp{-i{\cal H}t}\rangle 
+\langle \eexp{i{\cal H}t} \, V\rangle\langle V \, \eexp{-i{\cal H}t}\rangle 
\nonumber 
\\
\langle V(0)V(t)\rangle &=&
\langle V \, \eexp{i{\cal H}t}\rangle\langle V \, \eexp{-i{\cal H}t}\rangle 
+\langle V \, \eexp{i{\cal H}t} \, V\rangle\langle\eexp{-i{\cal H}t}\rangle 
\nonumber 
\\
\langle V(t)V(0)\rangle &=&
\langle\eexp{i{\cal H}t} \, V\rangle\langle\eexp{-i{\cal H}t} \, V\rangle 
+\langle\eexp{i{\cal H}t}\rangle\langle V \, \eexp{-i{\cal H}t} \, V\rangle
\nonumber 
\ee
It follows that 
\be{0}
C(0,0) &=& C(0) \\
C(t,t) &=& P_0(t)C(0) + \dot{c}_0(t)^2  \\ 
C(0,t) &=& C(t,0) \ \ = \ \ \dot{c}_0(t)^2 - c_0(t) \ddot{c}_0(t)
\ee
Substituting into Eq.(\ref{e46}) we get the result
\be{48} 
\Delta E_{\tbox{sprd}}(t) =
\Big[(1{+}{c_0}^2(t))C(0)-{\dot{c}_0}(t)^2+2{c_0}(t){\ddot{c}}_0(t)\Big]^{1/2}
\ \ \ \ 
\ee
We note that for short times ($t \ll t_0$) we can use the approximations
\be{0}
c_0(t) &\approx& 1\\
\dot{c}_0(t) &\approx& 0\\ 
\ddot{c}_0(t)&\approx& -C(t) 
\ee
which demonstrates the agreement with Eq.(\ref{e47}).
The numerical results in Fig.(\ref{fig:spreading}) 
confirm the validity of Eq.(\ref{e48}) for the FM,  
and Eq.(\ref{e47}) for the WM. 
We note that in the FM case the effect of recurrences
is more pronounced, because they are better synchronized:
all the out-in-out traffic goes exclusively through the initial state.
Fig.(\ref{fig:sat}) establishes the $\sqrt{2}$ ratio 
throughout the whole range of $s$~values. 
One should be aware that for small~$s$ 
the $(\omega_c^s/s)$ in Eq.(\ref{e477}) 
should be replaced by ${(\omega_c^s/s)-(\omega_{\varrho}^s/s)}$, 
which takes into account the {\em finite} 
level spacing in the numerical simulations. 
For very small~$s$ this goes to $\log(\varrho\omega_c)$,  
as if ${s=0}$. The finite level spacing effect 
clearly shows up in the numerics, 
and would not arise in the strict continuum limit.

\section{Summary and discussion}

In this paper, we have considered a quantum mechanical system, prepared in a discrete state, that subsequently 
decays into a non-Ohmic continuum of other states. Two different models that have the same spectral properties, but 
still different underlying dynamics have been analyzed and the respective results have been critically compared.
One model (FM) reflects integrable dynamics while the other is an RMT model (WM) that corresponds 
to chaotic dynamics. In both cases, a universal generalized Wigner time that governs the relaxation process
has emerged, while the non-universality is reflected in the decay law: 
We find that for ``non-Ohmic chaos" (WM case) the survival probability becomes a stretched exponential 
beyond the Wigner time scale, which is both surprising and interesting. This is contrasted 
with the ``integrable" power-law decay that takes over in the long time limit (FM case), 
and obviously is very different from the Ohmic exponential result. Only the standard case 
of Ohmic bandprofile is fully universal. 

We have also investigated the temporal behavior 
of the second moment of the spreading wavepacket. We have found that in the FM case 
the generalized Wigner time is reflected in the spreading process and not only
in the survival probability, contrary to the naive linear response theory expectation. 
 
Non-Ohmic coupling to the continuum emerges in various frameworks in physics.   
Quantized chaotic systems might exhibit non-Ohmic fluctuations due to semi-classically implied
long time power-law correlations, and in any case the typical power spectrum 
is in general not like ``white noise" (e.g. \cite{wls,lds,brm}).  
Other examples \cite{S01} appear in the context of a many-particle system, 
where the hierarchy of states and associated couplings, ranging from the single-particle levels 
to the exponentially dense spectrum of complicated many-particle excitations, can lead to a very 
structural non-Ohmic bandprofile describing the residual interactions. 
These non-universal structures of the bandprofile of the perturbation, 
lead to a highly non-linear decay of the survival probability. 
Depending on the context, the survival probability could be also 
related to the study of dephasing, or indirectly to the study 
of quantum fidelity and irreversibility: 
the generalized Wigner time is reinterpreted as the coherence time, 
in the same sense as in Landau's Fermi liquid theory.

It is worth mentioning, that in a bosonic second quantized
language the decay of the probability can be re-interpreted
as the decay of the site occupation~$\hat{n}$.
If the interaction between the bosons is neglected
this reduction is {\em exact} and merely requires an
appropriate dictionary.  In the latter context each level
becomes a bosonic site which is formally like an harmonic oscillator,
and hence the initially empty continuum is regarded as a zero temperature bath.
Consequently the decay problem is formally re-interpreted
as a {\em quantum dissipation} problem
with an Ohmic ($s{=}1$) or non-Ohmic ($s{\ne}1$) bath.
The generalized Wigner time scale is associated 
with the damped motion of the generalized coordinate~$\hat{n}$.

\begin{widetext}

\begin{table}[h]
\begin{tabular}{|c||c|c|c|}

\hline

model, $s$ & LDOS &  $P_0(t)$  \cr

\hline

FM, WM $s=1$ & Lorentzian &  Exponential decay \cr

\hline

WM, $0{<}s{<}2$ & Semicircle core + tails & Exponential-like decay\cr

\hline

FM, $0{<}s{<}2$ & Singular core + tails & Exponential-like followed by Power-law \cr

\hline

FM, $s=2$ & $1/\omega$ & $\mbox{Log}(t)$ decay \cr

\hline

FM, WM, $s>2$ & Delta core + tails & No decay after transient  \cr

\hline

$s\leq 0$ & Core + low weight tails & Infrared dominated \cr

\hline

\end{tabular}

\caption{The various results for the LDOS and for the survival probability at a glance.}

\end{table}

\end{widetext}

Table~1 summarizes the various results that we have obtained for 
the survival probability. We conclude this section with 
a somewhat technical discussion of the crossovers between the 
various $s$~regimes. Note that the statements below are implied by 
inspection of Fig.\ref{fig:stages}.  
The strictly universal Ohmic result holds for ${s=1}$. 
The super-Ohmic universal regime is bounded 
from above by ${s=2}$, but for finite ultraviolate cutoff~$\omega_c$ 
the effective border is a bit lower. This means that the marginally 
universal ${s=2}$ behavior prevails in a finite strip around ${s=2}$.
For ${s>2}$, if the coupling to the continuum is small enough,  
the survival probability does not decay, 
except a short transient that can be described by FOPT. 
Similarly, the sub-Ohmic universal regime is bounded 
from below by ${s=0}$, but for finite infrared cutoff the 
effective border is a bit higher. Below this $s$-border 
the FOPT tails of the LDOS become sub-dominant 
[as implied by the divergence of the first term in Eq.(\ref{e701})],  
and the decay of the survival probability becomes infrared determined:
This means that the effective cutoff is not~$\omega_c$,  
but some different ill-defined (model dependent) cutoff at much lower frequencies,
that might be determined by the level spacing statistics.

\appendix

\section{Green function formulation}

For convenience we take the energy reference as ${E_0=0}$.  
The Resolvent is defined as
\be{0}
G^{+}(\omega) = \frac{1}{\omega-{\cal H} + i0}
\ee
Substitution of ${\cal H}= {\cal H}_0 + V$, expansion 
to infinite order, and {\em exact} geometric summation, 
can be carried out in the FM case, leading to the  
following standard result:   
\be{0}
\langle 0|G^+(\omega)|0\rangle = \frac{1}{\omega- \Delta(\omega)+i(\Gamma(\omega)/2)}
\ee
Using the identity $\mbox{Im}[G^+]=-\pi\delta(\omega-{\cal H})$ 
one realizes that the LDOS is given by the expression 
\be{0}
\rho(\omega)=-\frac{1}{\pi} \mbox{Im}\left[ \langle 0|G^+(\omega)|0\rangle \right]
\ee
leading to Eq.(\ref{e20}). 
The evolution operator is given as the FT of the resolvent, 
namely ${U(t) = \mbox{FT}\left[-2\mbox{Im}[G^+(\omega)]\right]}$,    
hence the survival amplitude is 
\be{0}
c_0(t) = \langle 0|U(t)|0\rangle = \mbox{FT}\left[2\pi\rho(\omega)\right]
\ee 
in agreement with the elementary derivation in the text.

\section{Optional derivation of the survival amplitude formula}

In this appendix we give an optional direct derivation 
for the survival amplitude formula in the FM case without relaying 
on the theory of Green functions. 
We are interested in finding $c_n(t)$ and $c_0(t)$,
which are the amplitudes to find the particle in 
the respective levels. The Schrodinger equation is  
\be{112}
i\frac{\partial c_0}{\partial t} & = & \sum_{n}{\eexp{-i\omega_n t } V_{0,n} c_n(t)}
\\ \label{e113}
i\frac{\partial c_n}{\partial t} & = & \eexp{i\omega_n t } V_{n,0} c_0(t)
\ee
where $\omega_n \equiv E_n-E_0$.
By integration over (\ref{e113}) we get
\be{114}
c_n(t) = -\int_{0}^{t}\eexp{i\omega_n t'}V_{n,0}c_0(t')dt'
\ee
placing Eq.(\ref{e114}) into Eq.(\ref{e112}) we get 
\be{0}
\frac{dc_0}{dt} = -\int_{0}^{t}{C(t-t') \ c_0(t')\mbox{d}t'}
\ee
where 
\be{0}
C(\tau) = \mbox{FT}[\tilde{C}(\omega)]  = \sum_{n}|V_{n, 0}|^2\eexp{-i\omega_n \tau}
\ee

We want to solve the survival amplitude equation
using a Laplace transform technique. 
For that purpose we define $K(\tau) = \Theta(\tau)C(\tau)$, 
where $\Theta(\tau)$ is the Heaviside step function.
Then we rewrite the equations as 
\be{0}
\frac{dc_0}{dt} = \delta(t) -
\int_{-\infty}^{+\infty}K(t-t')c_0(t')\mbox{d}t'
\ee
where $c_0(\tau)$ is zero for negative $\tau$ and 
unknown for positive $\tau$. The corresponding 
equation for the Fourier components is 
\be{0}
-i\omega c_{\omega}=1-\widetilde{K}(\omega)c_{\omega}
\ee
From here it follows that the survival amplitude
can be written as an FT 
\be{0}
c_0(t) =  \int_{-\infty}^{+\infty}\frac{\mbox{d}\omega}{2\pi}
\left[\frac{\eexp{-i\omega t}}{-i\omega + \widetilde{K}(\omega)}\right]
\ee
From the definition of $K(\tau)$ and using the convolution 
theorem it follws that  
\be{0}
\widetilde{K}(\omega) = 
\frac{1}{2}\Gamma(\omega) + i\Delta(\omega)  
\ee
Hence consistency with Eq.(\ref{e317}) is established.

\section{The energy spreading formula}
\label{app:C}

For the derivation of the energy spreading formula
Eq.(\ref{e46}) it is convenient to
regard the Hamiltonian as time dependent,  
\be{0}
{\cal H}(\lambda(t)) = {\cal H}_0 + \lambda(t)V
\ee
where $\lambda(t)$ is a time dependent parameter.
We define generalized forces in the standard way
\be{0}
{\cal F} =  -\frac{\partial {\cal H}}
{\partial \lambda}(\lambda)
\ee
Using the Heisenberg picture,  
and the usual notation ${\cal F}(t)=U(t)^{-1}{\cal F}U(t)$, 
we have the following relation
\be{0}
\frac{\partial {\cal H}}
{\partial t} = -\dot{\lambda}(t){\cal F}(t)
\ee
Thus the change in the energy can be written as
\be{0}
{\cal H}(t)-{\cal H}(0) = -\int_0^t\dot{\lambda}(t'){\cal F}(t')dt'
\ee
Squaring and taking the expectation value with respect 
to the initial state we get  
\be{0}
\label{eD5}
\left[\Delta E_{\tbox{sprd}}(t)\right]^2 =
\int_0^tdt'\int_0^tdt''\dot{\lambda}(t')\dot{\lambda}(t'')C(t', \, t'')
\ee
where $C(t', t'')=\langle {\cal F}(t'){\cal F}(t'')\rangle$ 
is the auto-correlation function. 
The free evolution during the time interval ${[0,t]}$ 
corresponds formally to a ``rectangular pulse"
\be{0}
\lambda(t') = \Theta(t')-\Theta(t'-t)
\ee
where $\Theta()$ is a Heaviside step function. Its time derivative 
is ${\dot{\lambda} = \delta(t')-\delta(t'-t)}$, leading to Eq.(\ref{e46}).


\ \\ 
{\bf Acknowledgments:} 
This research is supported by the US-Israel Binational Science Foundation (BSF).
Some preliminary calculations have been done by Ori Ben-Dayan under the supervision of DC. 



\clearpage

\clearpage


\begin{thebibliography}{99}

\bibitem{AZ02}
N. Auerbach, V. Zelevinsky, Phys. Rev. C {\bf 65}, 034601 (2002);
V.V. Sokolov, V.G. Zelevinsky, Nucl. Phys. A {\bf 504}, 562 (1989).

\bibitem{CDG92}
C. Cohen-Tannoudji, J. Dupont-Roc, G. Grynberg,
{\it Atoms-Photon Interactions: Basic Processes and Applications} (Wiley, New-York, 1992).

\bibitem{NC00}
M.A. Nielsen and I.L. Chuang,
{\emph Quantum computation and quantum information}
(Cambridge University Press,2000).

\bibitem{PAEI94}
V.N. Prigodin, B.L. Altshuler, K.B. Efetov, S. Iida, Phys. Rev. Lett. {\bf 72}, 546 (1994);
B.L. Altshuler et al., Phys. Rev. Lett. {\bf 78}, 2803 (1997).

\bibitem{BH91}
C. W. J. Beenakker, H. van Houton,
in {\it Solid State Physics: Advances in Research and Applications},
Ed. H. Ehrenreich and D. Turnbull, 1-228 {\bf 44} (Academic Press, New York, 1991).

\bibitem{PRSB00}
E. Persson, I. Rotter, H.-J. St\"ockmann,
M. Barth, Phys. Rev. Lett. {\bf 85}, 2478 (2000).


\bibitem{WW30}
V. Weisskopf and E.P. Wigner,
Z. Phys. {\bf 63}, 54 (1930).

\bibitem{wigner}
E. Wigner,
Ann. Math {\bf 62} 548 (1955);
{\bf 65} 203 (1957).

\bibitem{mario} 
M. Feingold, Europhysics Letters {\bf 17}, 97 (1992).

\bibitem{flamb}
V.V. Flambaum, A.A. Gribakina, G.F. Gribakin and M.G. Kozlov, 
Phys. Rev. A {\bf 50}, 267 (1994).

\bibitem{izrailev}
F.M. Izrailev, A.Castaneda-Mendoza, Phys. Lett. A {\bf 350}, 355 (2006);
V.V. Flambaum, F.M.Izrailev, Phys. Rev. E {\bf 64} 026124 (2001);
V.V. Flambaum, F.M.Izrailev, Phys. Rev.E {\bf 61}, 2539 (2000).

\bibitem{fyodo}
Y.V. Fyodorov, O.A. Chubykalo, F.M. Izrailev, G. Casati, Phys. Rev. Lett. {\bf 76}, 1603 (1996).

\bibitem{GACMP97}
J.L. Gruver et al.,
Phys. Rev E {\bf 55}, 6370 (1997).


\bibitem{wls} 
D. Cohen, E.J. Heller, 
Phys. Rev. Lett. {\bf 84}, 2841 (2000).  

\bibitem{lds} 
D. Cohen and T. Kottos, 
Phys. Rev. E {\bf 63}, 36203 (2001).

\bibitem{brm} 
M. Hiller, D. Cohen, T. Geisel and T. Kottos,
Annals of Physics {\bf 321}, 1025 (2006).

\bibitem{HKG} 
M. Hiller, T. Kottos, T. Geisel, 
Phys. Rev. A {\bf 73}, 061604 (2006).



\bibitem{S01}
P.G. Silvestrov, Phys. Rev. B {\bf 64}, 113309 (2001); 
A. Amir, Y. Oreg, Y. Imry, Phys. Rev. A {\bf 77}, 050101(R) (2008).


\bibitem{haake}
F. Haake, ``Quantum Signatures of Chaos" (Springer 2000).

\bibitem{friedrichs}
K.O. Friedrichs,
Comm. Pure  Appl.  Math. \ {\bf 1}, 361 (1948).


\bibitem{wbr} 
D. Cohen, F.M. Izrailev, T. Kottos, 
Phys. Rev. Lett. {\bf 84}, 2052 (2000).

\bibitem{ref20} 
T. Kottos, D. Cohen,
Europhysics Letters {\bf 61}, 431 (2003).

\bibitem{frc} 
D. Cohen, Annals of Physics {\bf 283}, 175 (2000). 



\end{thebibliography}
\end{document}